# Dual Solutions for Opposing Mixed Convection in Porous Media

## Jian-Jun SHU[*,a], Qi-Wen WANG[b] and Ioan POP[c]

[a] *School of Mechanical & Aerospace Engineering, Nanyang Technological University, 50 Nanyang Avenue, Singapore 639798*

[b] *School of Business, Shanghai DianJi University, 1350 Ganlan Road, Lingang New City, Pudong New District, Shanghai 201306, People's Republic of China*

[c] *Department of Mathematics, Babeş-Bolyai University, 400084 Cluj-Napoca, Romania*

## ABSTRACT

*The problem of steady mixed convection boundary-layer flow on a cooled vertical permeable circular cylinder embedded in a fluid-saturated porous medium is studied. Here, we evaluate the flow and heat transfer characteristics numerically for various values of the governing parameters and demonstrate the existence of dual solutions beyond a critical point.*

*Keywords: Mixed convection, porous medium, vertical permeable circular cylinder, dual solutions*

## 1 Introduction

Convective heat transfer in fluid-saturated porous media has received a great amount of attention during the last few decades. This has been driven by its importance in many aspects of natural and industrial problems, such as the utilization of geothermal energy, chemical engineering, food processing and storage, nuclear waste management, thermal insulation system, contaminant transport in ground water, migration of moisture through air contained in fibrous insulation, and many others. Several reviews of the subject of convective flow in porous media were done by various researchers [1-6].

Mixed convection from a vertical cylinder embedded in a porous medium is the principal mode of heat transfer in numerous applications such as in connection with oil/gas lines, insulation of vertical porous pipes, cryogenics as well as in the context of water distribution lines, underground electrical power transmission lines, and disposal of radioactive waste, to name just a few applications. The case of free and mixed convection flow from a vertical cylinder placed in a porous medium has been studied extensively both analytically and numerically. A numerical solution of the problem of free convective boundary-layer flow induced by a heated vertical cylinder embedded in a fluid-saturated porous medium was presented by Minkowycz and Cheng [7] when the surface temperature of cylinder was taken to be proportional to $x^n$ where $x$ was the distance from the leading edge of cylinder and $n$ was a constant. The results were obtained for various values of $n$ lying between $0$ and $1$, by using similarity and local nonsimilarity methods [8-14]. The problem was later extended by various researchers [15-22]. Bassom and Rees [19] studied the free convection boundary-layer flow induced by a heated vertical cylinder which was embedded in a fluid-saturated porous medium, with the surface temperature of the cylinder varying as $x^n$. Both numerical and asymptotic analyses were presented for the governing nonsimilar boundary-layer equations. When $n < 1$, the asymptotic flow field far from the leading edge of cylinder was taken on a multiple-layer structure. On the other hand, for $n > 1$, only a simple single layer was present far downstream, but a multiple layer structure existed close to the leading edge of the cylinder.

---

[*] Correspondence should be addressed to Jian-Jun SHU, mjjshu@ntu.edu.sg

In this paper, we consider the problem of mixed convection boundary-layer flow along a cooled vertical permeable circular cylinder embedded in a fluid-saturated porous medium, shown in Fig. 1. It is assumed that the mainstream velocity $U(x)$ and surface temperature $T_w(x)$ of the cylinder vary linearly with the distance $x$ along the cylinder. It is also assumed that the axially symmetric surface mass flux $\omega$ is constant. The similarity equation involves three parameters, namely, the buoyancy convection parameter $\lambda$, curvature parameter $\gamma$, and suction or blowing parameter $\sigma$. It should be stated at this end that mixed convection flows, or the combination of both free and forced convection, occur in many transport processes in natural and industrial applications, including electronic device cooled by a fan, nuclear reactor during emergency shutdown, heat exchange in low-velocity environment, and solar receiver exposed to wind current. The effect of buoyancy induced flow in forced convection or forced flow in free convection becomes significant for such transport processes. When the flow velocity is relatively low and the temperature difference between the surface and the free stream is relatively large, thermal buoyancy forces play a significant role in forced convection heat transfer as well as in the onset of flow instabilities, because of being responsible for delaying or speeding up the transition from laminar to turbulent flow (see Chen and Armaly [23]). These authors have shown that the mixed convection regime is $a \leq \dfrac{G_r}{R_e^{\,n}} \leq b$, where $G_r$ is the Grashof number, $R_e$ is the Reynolds number, $n$ is a constant, which depends on flow configuration and surface heating condition, and $a$ and $b$ are the lower and upper bounds of regime, respectively. The buoyancy parameter $\dfrac{G_r}{R_e^{\,n}}$ represents a measure of the effect of free convection in comparison to that of forced convection on the flow. Outside the mixed convection regime, the analysis of a pure either free or forced convection can be adopted to describe the flow and temperature field accurately.

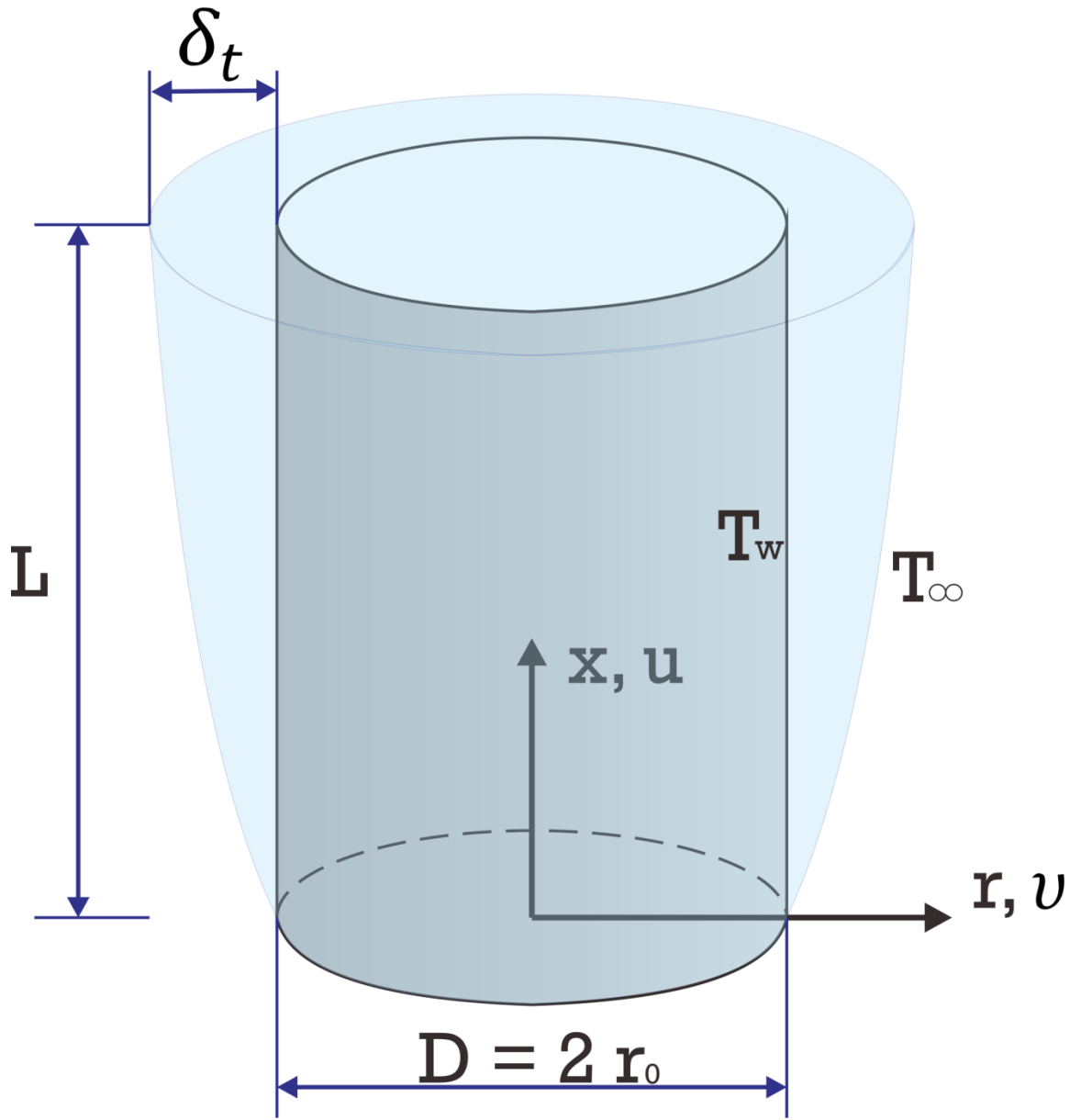

**Fig. 1** Physical model and coordinate system

## 2 Governing Equations

For the Darcy steady mixed convection flow of a viscous incompressible fluid along the vertical permeable circular cylinder of radius $r_0$ embedded in a fluid-saturated porous medium with prescribed axially symmetric velocity $v_w$, wall temperature $T_w(x)$ and mainstream velocity $U(x)$ in fluid at constant ambient temperature $T_\infty$, the governing equations for continuity, Darcy with Boussinesq approximation, and energy can be written by using the usual boundary-layer approximation as (see Nield and Bejan
[5], Merkin and Pop [17] or Bassom and Rees [19])

$$\frac{\partial}{\partial x}(r\,u) + \frac{\partial}{\partial r}(r\,v) = 0 \ , \tag{1}$$

$$u = U(x) + \frac{g\,\beta\,K}{\nu}(T - T_\infty) \ , \tag{2}$$

$$u\frac{\partial T}{\partial x} + v\frac{\partial T}{\partial r} = \alpha_m\left(\frac{\partial^2 T}{\partial r^2} + \frac{1}{r}\frac{\partial T}{\partial r}\right) \ , \tag{3}$$

subject to the boundary conditions

$$v = v_w, \qquad T = T_w(x) \qquad \text{at} \qquad r = r_0 \ , \tag{4}$$

$$u \rightarrow U(x), \qquad T \rightarrow T_\infty \qquad \text{as} \qquad r \rightarrow \infty \ . \tag{5}$$

Here, the coordinates $x$ and $r$ measure distance along the surface and normal to it,
respectively; $u$ and $v$ are the velocity components along $x$ and $r$ axes; $T$ is the fluid
temperature; $g$ is the acceleration due to gravity; $K$ is the permeability of the porous
medium; $\nu$ is the kinematic viscosity; $\alpha_m$ is the effective thermal diffusivity; and $v_w$ is
the velocity of suction ($v_w < 0$) or blowing ($v_w > 0$), respectively. Following Mahmood
and Merkin [18], we assume in this paper that

$$U(x) = \frac{U_\infty\,x}{L} \qquad \text{and} \qquad T_w(x) = T_\infty + \frac{x\,\Delta T}{L} \ , \tag{6}$$

where $\Delta T$ and $L$ are the temperature and length characteristics. With this choice of
mainstream and cylinder temperature, Eqs. (1)-(3) can be reduced to similarity form by
introducing the variables

$$\psi = \frac{2\,\alpha_m\,x}{\gamma}f(\eta), \quad T = T_\infty + \frac{x\,\Delta T}{L}\theta(\eta), \quad \eta = \frac{r^2 - r_0^{\,2}}{r_0^{\,2}\,\gamma} \ , \tag{7}$$

where $\gamma = \frac{2}{r_0}\sqrt{\dfrac{\alpha_m\,L}{U_\infty}}$ is the curvature parameter and $\psi$ is the stream function defined in

the usual way

$$u = \frac{1}{r}\frac{\partial\psi}{\partial r} \quad \text{and} \quad v = -\frac{1}{r}\frac{\partial\psi}{\partial x} \ . \tag{8}$$

Eqs. (1)-(3) then become

$$f' = 1 + \lambda\,\theta \ , \tag{9}$$

$$(1 + \gamma\,\eta)\theta'' + \gamma\,\theta' + f\,\theta' - f'\,\theta = 0 \ , \tag{10}$$

together with the boundary conditions

$$f(0) = \sigma, \quad \theta(0) = 1, \quad \theta(\infty) = 0 \ . \tag{11}$$

Here, $\lambda = \dfrac{Ra}{Pe}$ is the buoyancy parameter (ratio of free to forced convection velocity scales), $Ra = \dfrac{g\,\beta\,K\,L\,\Delta T}{\alpha_m\,\nu}$ is the Rayleigh number, $Pe = \dfrac{U_\infty L}{\alpha_m}$ is the Péclet number, and

$\sigma = -\dfrac{v_w\,r_0\,\gamma}{2\,\alpha_m}$ is the suction ($\sigma > 0$) or blowing ($\sigma < 0$) parameter. Primes denote

differentiation with respect to $\eta$. Combining (9) and (10), we finally have the equation

$$(1 + \gamma\,\eta)f''' + f\,f'' + \gamma\,f'' + f' - f'^2 = 0 \ , \tag{12}$$

along with the boundary condition

$$f(0) = \sigma, \quad f'(0) = 1 + \lambda, \quad f'(\infty) = 1 \ . \tag{13}$$

The physical quantity of interest is the skin friction coefficient $C_f$, which is

defined as

$$C_f = \frac{\tau_w}{\dfrac{1}{2}\rho U_\infty U} \ , \tag{14}$$

where the skin friction $\tau_w$ is given by

$$\tau_w = \mu\left(\frac{\partial u}{\partial r}\right)_{r=r_0} \ . \tag{15}$$

Using the similarity variables (7), we have the reduced skin friction

$$f''(0) = \frac{C_f\,\alpha_m\,\sqrt{Pe}}{2\,\nu} \ . \tag{16}$$

## 3  Results and Discussion

Eq. (12) with the boundary conditions (13) has been solved numerically for the selected values of the governing parameters by using the standard numerical method [24].

Our principal objectives being to assess the effects that mixed convection, curvature, and suction parameters have on the flow and heat transfer characteristics. It is worth pointing out that some special cases have been considered for $\gamma\left[\gamma(2+\lambda)+\sigma\right]=1$ [25], $\gamma = 0, \sigma = 0$ [26], and $\sigma = 0$ [27]. All values of three parameters, the buoyancy convection parameter $\lambda$, curvature parameter $\gamma$, and suction or blowing parameter $\sigma$, are presented as follows.

In Figs. 2 to 4, the reduced skin friction $f''(0)$ is plotted against the mixed convection parameter $\lambda$ for various values of the curvature parameter $\gamma$ and mass flux parameter $\sigma$ for the case of opposing flow ($\lambda < 0$). All these figures show that dual solutions (upper and lower branch solutions) exist for Eq. (12) with the boundary conditions (13) for all values of $\lambda_c \le \lambda < 0$, where $\lambda_c < 0$ is the critical value of $\lambda < 0$, and all values of $\gamma$ and $\sigma$ considered. The two branches (upper and lower branch solutions) merge with one another at a critical point $\lambda_c < 0$, where the boundary-layer solutions beyond this point are impossible to be obtained due to boundary-layer separation from the surface. The full Darcy and energy equations should be solved for $\lambda_c \le \lambda < 0$. Both solution branches are passing through the forced convection solution $f = \sigma + \eta$ at $\lambda = 0$ without a singularity appearing, as seen, for example, in [25]. Even though a solution to (12, 13) exists on the lower branch when $\lambda = 0$, it cannot be a physically acceptable (realizable) solution to our original problem. Figs. 2 to 4 also show that when the curvature is increased from $\gamma = 0$ (flat plate) to $\gamma = 10$, it increases the range of existence of solutions and the reduced skin friction $f''(0)$. In addition, Fig. 4 illustrates that when the mass flux parameter $\sigma$ is decreased from $\sigma = 0$ (impermeable cylinder) to $\sigma = -8.49$ (injection), it increases the range of existence of solutions and the reduced skin friction $f''(0)$. A stability analysis by adopting the techniques of Merkin [16] and Wilks and Bramley [28] reveals that the upper branch solutions are stable and physically realizable, while the lower branch solutions are unstable and, therefore, not physically realizable. It is to be noticed that the problem (12, 13) admits an exact solution for the special case $\gamma\left[\gamma(2+\lambda)+\sigma\right]=1$ (see Magyari *et al.* [25])

$$f(\eta) = \sigma + \eta + \lambda\gamma\left(1 - e^{-\frac{\eta}{\gamma}}\right).$$ (17)

It should be stated that the results obtained by Magyari *et al*. [25] are, in fact, identical with the numerical results obtained from Eq. (12) subject to (13) in this paper. We found, however, that our numerical results are quantitatively consistent with the analytical results reported by Magyari *et al*. [25]. Thus, it gives us confidence that the present numerical results are correct for all values of $\lambda$, $\gamma$, and $\sigma$ considered.

Finally, we also include the plots of the velocity profile $f'(\eta)$ in Fig. 5, which show the existence of dual solutions of the problem (12, 13). It is clearly seen from these figures that boundary-layer thickness becomes thinner for the upper branch solution as compared to the lower branch solution and the far-field boundary conditions (13) are satisfied asymptotically. Therefore, it confirms the validity of the numerical results and the existence of the dual solutions illustrated in Figs. 2 and 4.

## 4 Conclusion

In summary, the problem of steady mixed convection boundary-layer flow on a cooled vertical permeable circular cylinder embedded in a fluid-saturated porous medium is studied. We take particular forms for the outer flow and wall temperature variation that enable the system of the partial differential equations to be reduced to a similarity form, Eqs. (12, 13). For an opposing flow ($\lambda < 0$), we find that there is a critical value $\lambda_c < 0$ of the mixed convection parameter $\lambda < 0$, with solutions existing only for $\lambda_c \leq \lambda \leq 0$. However, for $\lambda \leq \lambda_c < 0$, the solutions of the problem (12, 13) do not exist. We then examined the effects of the curvature $\gamma$ and mass flux $\sigma$ parameters on the reduced skin friction $f''(0)$ and velocity profile $f'(\eta)$. Graphical qualitative comparison has been made with the existing results in literature and it is found to be in good agreement. It is worth mentioning at this end that Wilks and Bramley [28] presented dual similarity solutions in the context of mixed convection flow. They showed that for this flow, dual solution existed and they displayed reverse flow. In contrast to the Falkner-Skan solutions, the bifurcation point was to be distinct from the point of vanishing skin friction. A significant feature of the new solutions discovered by Wilks and Bramley [28] was the location of the bifurcation point, separating two branches of solution, away from the point of vanishing skin friction.

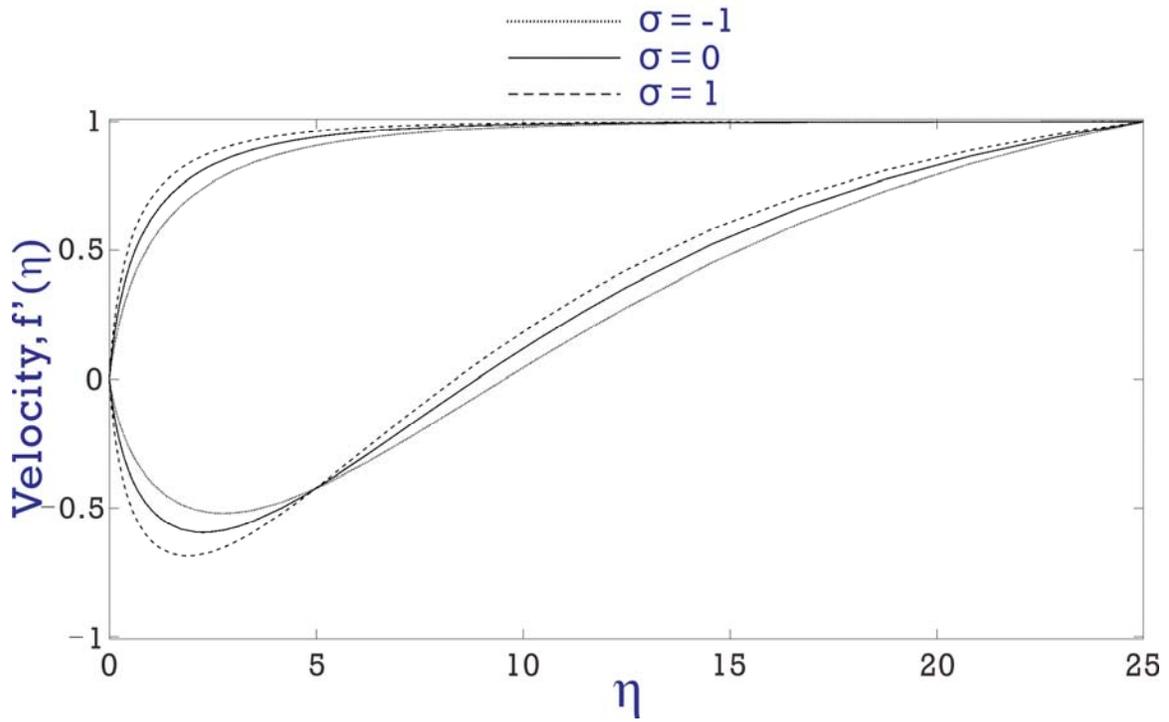

**Fig. 2** Velocity profile $f'(\eta)$ for various $\sigma$ at $\lambda = -1$ and $\gamma = 5$

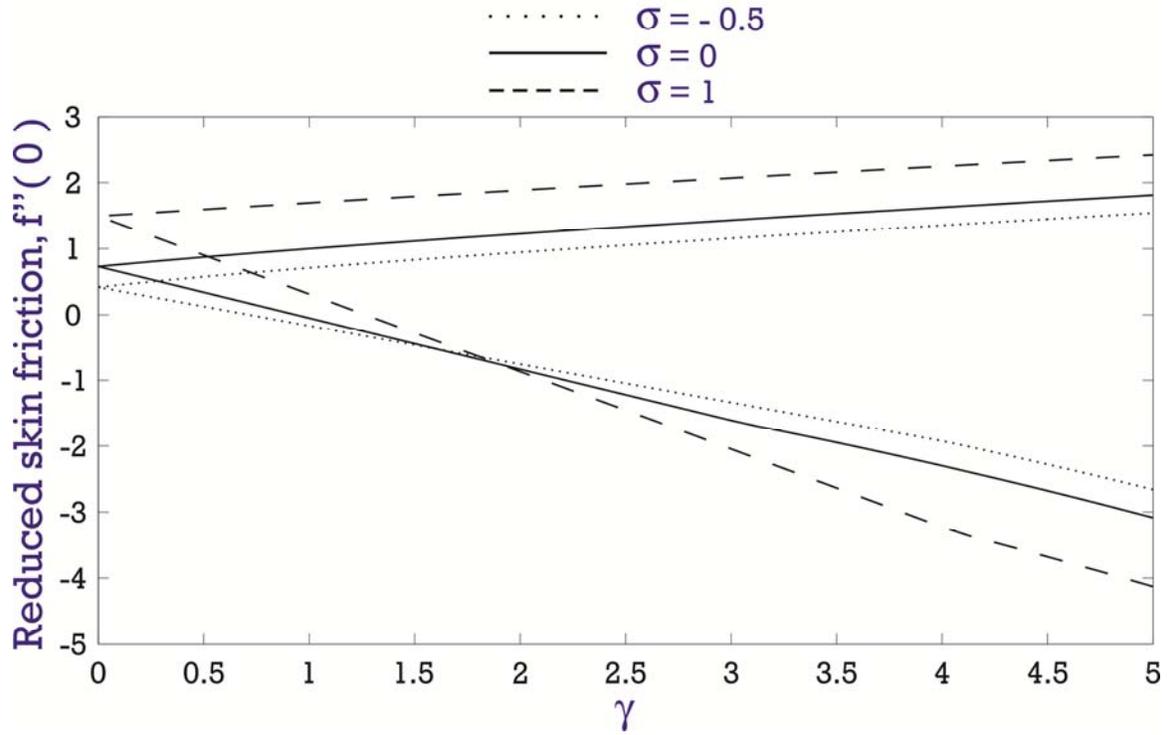

**Fig. 3** Reduced skin friction $f''(0)$ with $\gamma$ for various $\sigma$ at $\lambda = -1$

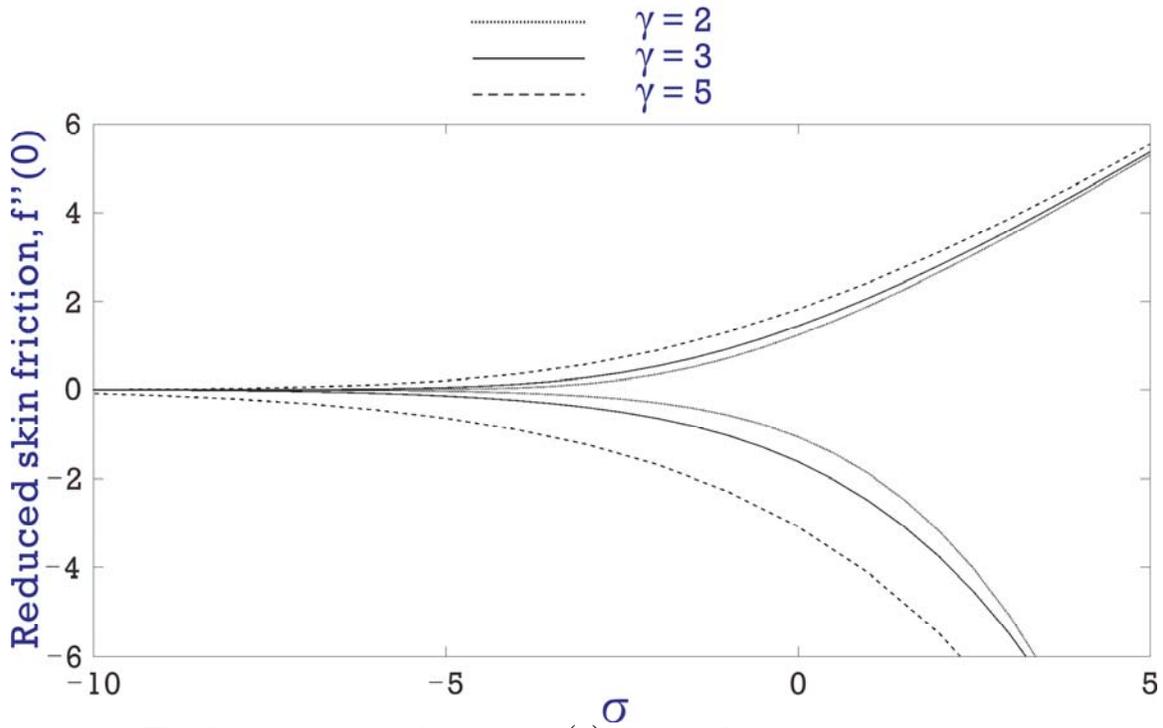

**Fig. 4** Reduced skin friction $f''(0)$ with $\sigma$ for various $\gamma$ at $\lambda = -1$

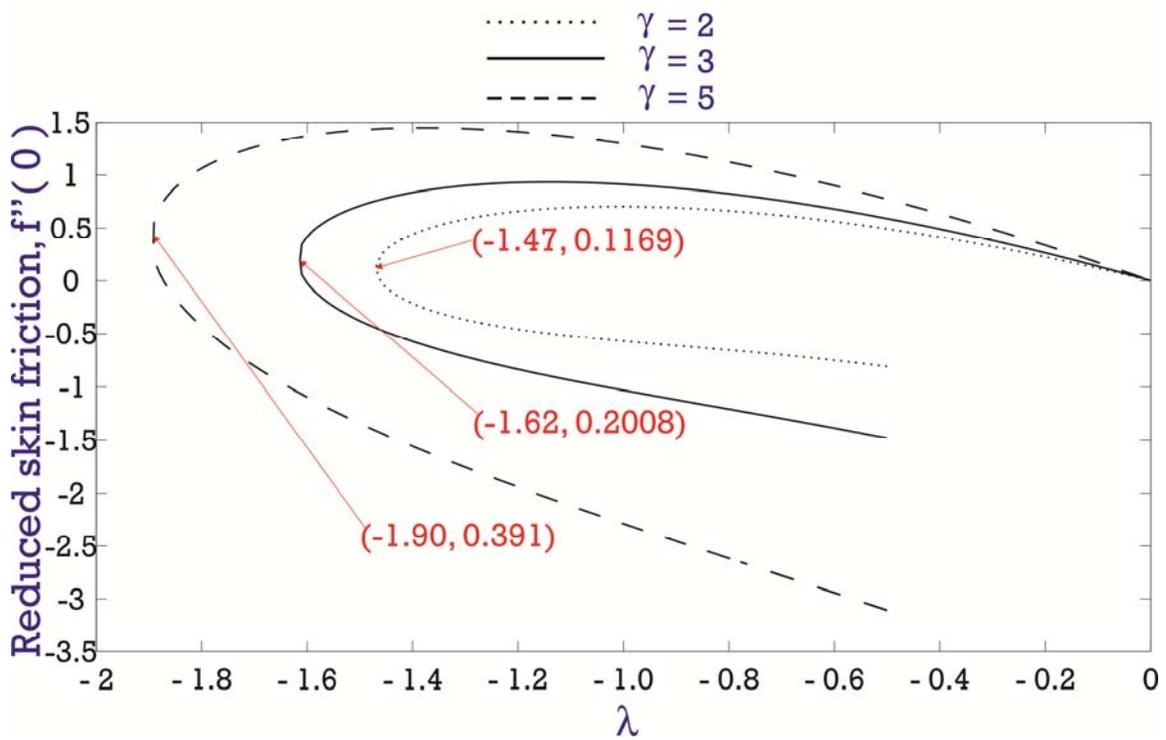

**(a)**

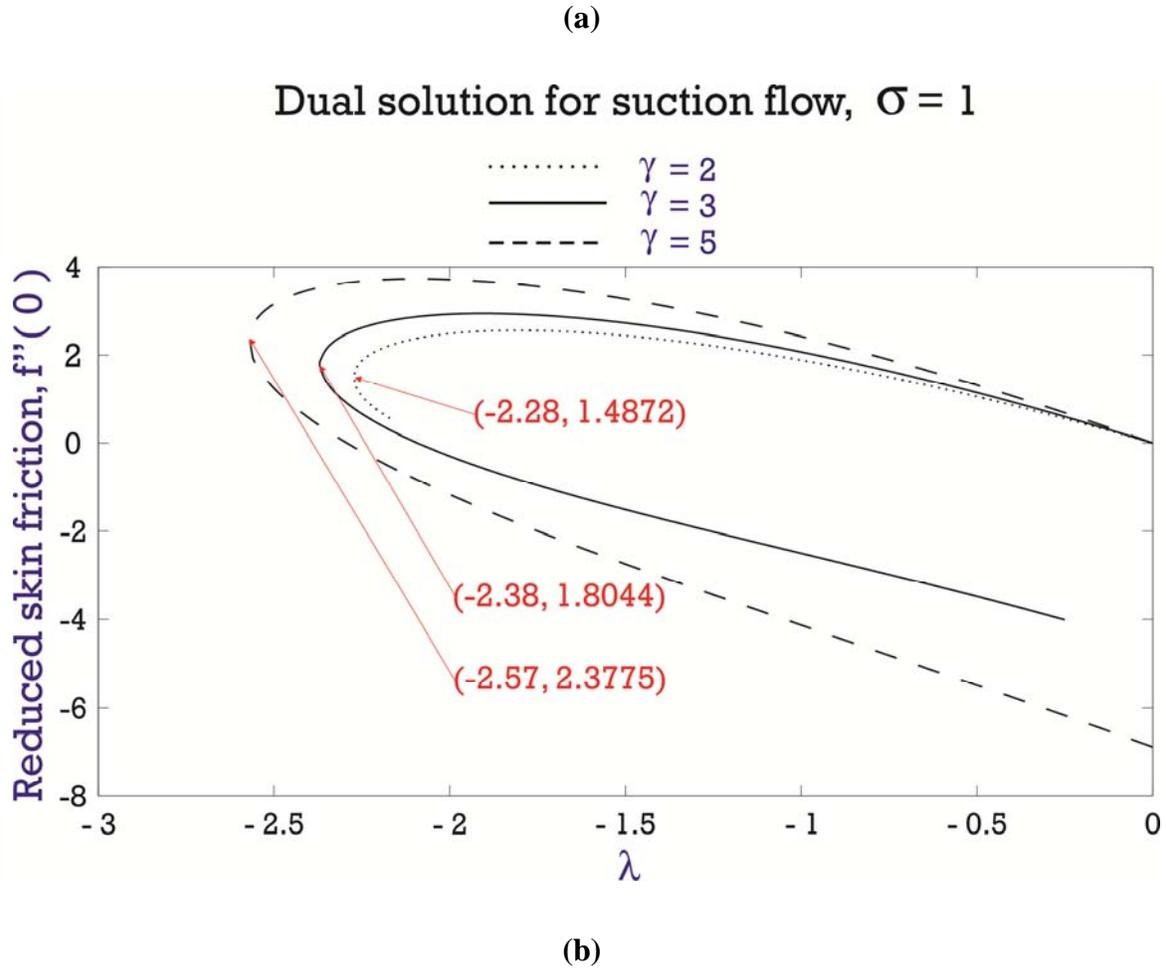

**(b)**

**Fig. 5** Reduced skin friction $f''(0)$ with $\lambda$ for various $\gamma$

## Acknowledgements

This work was supported by Nanyang Technological University (M4081942).

## Nomenclature

| | |
|---|---|
| $f$ | dimensionless stream function |
| $g$ | gravitational acceleration |
| $K$ | permeability of a porous medium |
| $L$ | length characteristic |
| $P_e$ | Péclet number for a porous medium |
| $r$ | radial coordinate |
| $r_0$ | radius of cylinder |
| $R_a$ | Rayleigh number for a porous medium |
| $T$ | fluid temperature |
| $u$ , $v$ | velocity components in $x$ - and $r$ -directions |
| $U(x)$ | mainstream velocity in axial direction |
| $x$ | axial coordinate |

### Greek Symbols

| | |
|---|---|
| $\alpha_m$ | equivalent thermal diffusivity |
| $\beta$ | coefficient of thermal expansion |
| $\gamma$ | curvature parameter |
| $\Delta T$ | characteristic temperature |
| $\eta$ | pseudosimilarity variable |
| $\theta$ | dimensionless temperature |
| $\lambda$ | mixed convection parameter |
| $\nu$ | kinematic viscosity |
| $\sigma$ | suction or injection parameter |
| $\psi$ | stream function |

### Subscripts

| | |
|---|---|
| $w$ | condition at wall |
| $\infty$ | condition in ambient fluid |

### Superscript

| | |
|---|---|
| , | differentiation with respect to $\eta$ |

### References

[1]   Pop, I., and Ingham, D. B., 2001, *Convective Heat Transfer: Mathematical and Computational Modelling of Viscous Fluids and Porous Media*, Pergamon, Oxford.

[2]   Ingham, D. B., and Pop, I., 2005, *Transport Phenomena in Porous Media*, Elsevier Science, Oxford.

[3]   Vadász, P., 2008, *Emerging Topics in Heat and Mass Transfer in Porous Media: From Bioengineering and Microelectronics to Nanotechnology*, Springer Verlag, New York.

[4]   Vafai, K., 2010, *Porous Media: Applications in Biological Systems and Biotechnology*, CRC Press, Tokyo.

[5]   Nield, D. A., and Bejan, A., 2013, *Convection in Porous Media* (4th edition), Springer, New York.

[6]   Vafai, K., 2014, *Handbook of Porous Media* (3rd edition), CRC Press.

[7]   Minkowycz, W. J., and Cheng, P., 1976, "Free convection about a vertical cylinder embedded in a porous medium," *International Journal of Heat and Mass Transfer*, **19**(7), pp. 805–813.

[8]   Shu, J.-J., and Wilks, G., 1995, "Mixed-convection laminar film condensation on a semi-infinite vertical plate," *Journal of Fluid Mechanics*, **300**, pp. 207–229.

[9]   Shu, J.-J., and Wilks, G., 1996, "Heat transfer in the flow of a cold, two-dimensional vertical liquid jet against a hot, horizontal plate," *International Journal of Heat and Mass Transfer*, **39**(16), pp. 3367–3379.

[10]  Shu, J.-J., and Pop, I., 1997, "Inclined wall plumes in porous media," *Fluid Dynamics Research*, **21**(4), pp. 303–317.

[11]  Shu, J.-J., and Pop, I., 1998, "On thermal boundary layers on a flat plate subjected to a variable heat flux," *International Journal of Heat and Fluid Flow*, **19**(1), pp. 79–84.

[12]  Shu, J.-J., and Pop, I., 1999, "Thermal interaction between free convection and forced convection along a vertical conducting wall," *Heat and Mass Transfer*, **35**(1), pp. 33–38.

[13]  Shu, J.-J., and Wilks, G., 2008, "Heat transfer in the flow of a cold, axisymmetric vertical liquid jet against a hot, horizontal plate," *Journal of Heat Transfer-Transactions of the ASME*, **130**(1), pp. 012202.

[14]  Shu, J.-J., 2012, "Laminar film condensation heat transfer on a vertical, non-isothermal, semi-infinite plate," *Arabian Journal for Science and Engineering*, **37**(6), pp. 1711–1721.

[15] Merkin, J. H., 1980, "Mixed convection boundary-layer flow on a vertical surface in a saturated porous-medium," *Journal of Engineering Mathematics*, **14**(4), pp. 301–313.

[16] Merkin, J. H., 1986, "On dual solutions occurring in mixed convection in a porous-medium," *Journal of Engineering Mathematics*, **20**(2), pp. 171–179.

[17] Merkin, J. H., and Pop, I., 1987, "Mixed convection boundary-layer on a vertical cylinder embedded in a saturated porous medium," *Acta Mechanica*, **66**(1-4), pp. 251–262.

[18] Mahmood, T., and Merkin, J. H., 1988, "Similarity solutions in axisymmetric mixed-convection boundary-layer flow," *Journal of Engineering Mathematics*, **22**(1), pp. 73–92.

[19] Bassom, A. P., and Rees, D. A. S., 1996, "Free convection from a heated vertical cylinder embedded in a fluid-saturated porous medium," *Acta Mechanica*, **116**(1-4), pp. 139–151.

[20] Shu, J.-J., and Pop, I., 1998, "Transient conjugate free convection from a vertical flat plate in a porous medium subjected to a sudden change in surface heat flux," *International Journal of Engineering Science*, **36**(2), pp. 207–214.

[21] Shu, J.-J., and Wilks, G., 2009, "Heat transfer in the flow of a cold, two-dimensional draining sheet over a hot, horizontal cylinder," *European Journal of Mechanics B-Fluids*, **28**(1), pp. 185–190.

[22] Shu, J.-J., and Wilks, G., 2013, "Heat transfer in the flow of a cold, axisymmetric jet over a hot sphere," *Journal of Heat Transfer-Transactions of the ASME*, **135**(3), pp. 032201.

[23] Chen, T. S., and Armaly, B. F., 1987, *Mixed convection in external flow*. In: Handbook of Single-Phase Convective Heat Transfer (S. Kakaç, R.K. Shah, W. Aung, eds.). Wiley, New York, 14.1–14.35.

[24] Shu, J.-J., and Wilks, G., 1995, "An accurate numerical method for systems of differentio-integral equations associated with multiphase flow," *Computers & Fluids*, **24**(6), pp. 625–652.

[25] Magyari, E., Pop, I., and Keller, B., 2005, "Exact solutions for a longitudinal steady mixed convection flow over a permeable vertical thin cylinder in a porous

medium," *International Journal of Heat and Mass Transfer*, **48**(16), pp. 3435–3442.

[26] Merrill, K., Beauchesne, M., Previte, J., Paullet, J., and Weidman, P., 2006, "Final steady flow near a stagnation point on a vertical surface in a porous medium," *International Journal of Heat and Mass Transfer*, **49**(23-24), pp. 4681–4686.

[27] Rohni, A. M., Ahmad, S., Merkin, J. H., and Pop, I., 2013, "Mixed convection boundary-layer flow along a vertical cylinder embedded in a porous medium filled by a nanofluid," *Transport in Porous Media*, **96**(2), pp. 237–253.

[28] Wilks, G., and Bramley, J. S., 1981, "Dual solutions in mixed convection," *Proceedings of the Royal Society of Edinburgh Section A-Mathematics*, **87**(3-4), pp. 349–358.